%% file: example_paper.tex
\documentclass{article}

\usepackage{microtype}
\usepackage{graphicx}
\usepackage{subcaption}
\usepackage{booktabs} % for professional tables

\usepackage{hyperref}

\usepackage[preprint]{icml2026}

\usepackage{amsmath}
\usepackage{amssymb}
\usepackage{mathtools}
\usepackage{amsthm}

\usepackage[capitalize,noabbrev]{cleveref}

\theoremstyle{plain}

\theoremstyle{definition}

\theoremstyle{remark}

\usepackage[textsize=tiny]{todonotes}
\usepackage{tikz}
\usepackage{pgfplots}
\usepackage{siunitx}
\usepackage{enumitem}

\icmltitlerunning{SpooFL: Spoofing Federated Learning}

\begin{document}

\twocolumn[
  \icmltitle{SpooFL: Spoofing Federated Learning}



  \icmlsetsymbol{equal}{*}

  \begin{icmlauthorlist}
    \icmlauthor{Isaac Baglin}{yyy}
    \icmlauthor{Xiatian Zhu}{yyy}
    \icmlauthor{Simon Hadfield}{yyy}
  \end{icmlauthorlist}

  \icmlaffiliation{yyy}{CVSSP, University of Surrey, Guildford, United Kingdom}

\icmlcorrespondingauthor{Isaac Baglin}{ib00304@surrey.ac.uk}
\icmlcorrespondingauthor{Xiatian Zhu}{xiatian.zhu@surrey.ac.uk}
\icmlcorrespondingauthor{Simon Hadfield}{s.hadfield@surrey.ac.uk}
  \icmlkeywords{Machine Learning, ICML}

  \vskip 0.3in
]



\printAffiliationsAndNotice{}  % no special notice (required even if empty)

\begin{abstract}
\vspace{-3pt}
Traditional defenses against Deep Leakage (DL) attacks in Federated Learning (FL) primarily focus on obfuscation, introducing noise, transformations or encryption to degrade an attacker's ability to reconstruct private data. While effective to some extent, these methods often still leak high-level information such as class distributions or feature representations, and are frequently broken by increasingly powerful denoising attacks. We propose a fundamentally different perspective on FL defense: framing it as a spoofing problem.We introduce SpooFL (Figure \ref{fig:spoofl}), a spoofing-based defense that deceives attackers into believing they have recovered the true training data, while actually providing convincing but entirely synthetic samples from an unrelated task. Unlike prior synthetic-data defenses that share classes or distributions with the private data and thus still leak semantic information, SpooFL uses a state-of-the-art generative model trained on an external dataset with no class overlap. As a result, attackers are misled into recovering plausible yet completely irrelevant samples, preventing meaningful data leakage while preserving FL training integrity. We implement the first example of such a spoofing defense, and evaluate our method against state-of-the-art DL defenses and demonstrate that it successfully misdirects attackers without compromising model performance significantly.
\end{abstract}

\input{sec/1_intro}

\input{sec/2_lit}
\input{sec/2.5_spoof}
\input{sec/3_method}

\input{sec/4_eval}
\input{sec/5_conc}

\vspace{-5pt}
%\section*{Impact Statement}
\vspace{-5pt}
%This paper presents work whose goal is to advance the field of privacy-preserving machine learning by improving defenses against data reconstruction attacks in federated learning. The techniques studied here aim to reduce unintended leakage of sensitive training data while maintaining model utility. We do not anticipate any significant negative societal consequences beyond those commonly associated with research on machine learning security and privacy.

\bibliography{example_paper}
\bibliographystyle{icml2026}

\end{document}

%% file: sec/1_intro.tex
\vspace{-30pt}
\section{Introduction}
\label{sec:intro}
\vspace{-5pt}
Federated Learning (FL) is a decentralized approach to machine learning where multiple clients collaboratively train a shared model without exchanging raw data. Instead of sending private datasets to a central server, each client computes local model updates (typically gradients) which are then aggregated to improve the global model \cite{mcmahan2023communicationefficientlearningdeepnetworks}. This allows participants to jointly train a model using data that they cannot or do not want to release.
\begin{figure}[H]
    \centering
    \includegraphics[width=1\linewidth]{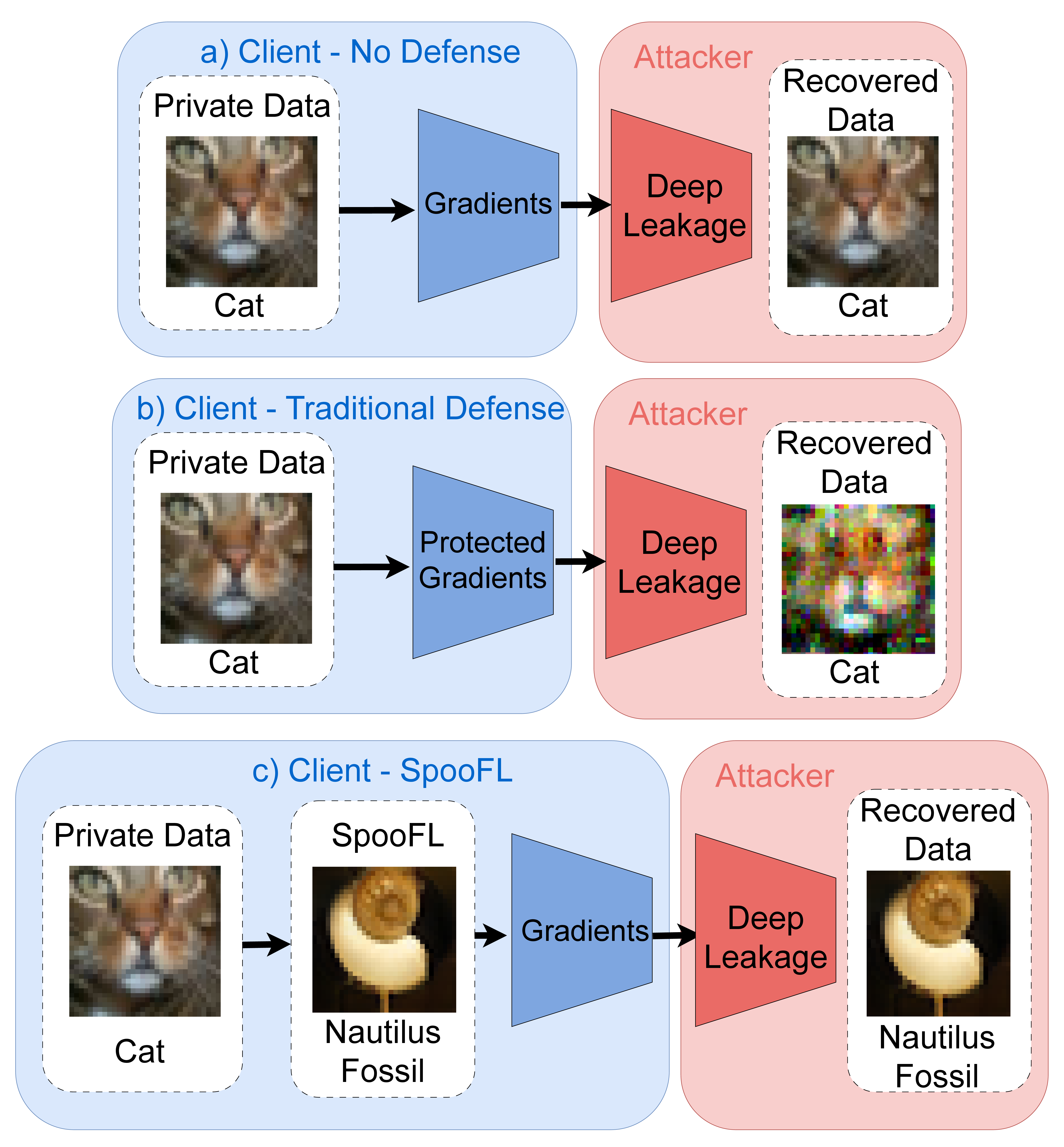}
    \caption{Gradient leakage under different defenses: a) No Defense yields near-exact reconstructions, b) Traditional Defenses obscure but retain features, c) SpooFL produces unrelated images, preserving privacy.}
    \label{fig:spoofl}
\end{figure}
\vspace{-15pt}
The two most common FL optimization strategies are Federated Stochastic Gradient Descent (FedSGD) and Federated Averaging (FedAvg). In FedSGD, clients compute gradients on their local data and send them to the server in every iteration, enabling frequent updates but increasing communication overhead. In contrast, FedAvg performs multiple local training steps before averaging the model weights across clients, reducing communication but potentially leading to stale updates \cite{rodio2024fedstaleleveragingstaleclient, karimireddy2021scaffoldstochasticcontrolledaveraging}. Despite FL’s decentralized nature, both methods remain vulnerable to gradient-based privacy attacks, such as Deep Leakage (DL), where adversaries exploit shared gradients to reconstruct private training samples \cite{zhu2019deepleakagegradients}. FedSGD is particularly vulnerable since it transmits raw gradients at every iteration, while FedAvg provides some natural obfuscation by averaging updates over multiple local steps \cite{geiping2020invertinggradientseasy}. However, even with FedAvg, an attacker can still extract meaningful information, especially if the batch size is small \cite{yin2021gradientsimagebatchrecovery}. DL attacks work by optimizing a randomly initialized image to minimize the difference between its computed gradients and the shared gradients from the client. We consider two potential threat models: (1) an honest-but-curious server that follows the FL protocol but inspects gradients or model updates to infer private information, and (2) a man-in-the-middle attacker who passively intercepts FL communications \cite{zhao2022deepleakagemodelfederated}. In both cases, the attacker has access to model weights and/or gradients but may lack private training data, task metadata, and the semantics of class labels, making reconstruction attacks an attractive strategy to recover this missing information. However, the attacker can iteratively refine an image until the generated gradients match those submitted by a client. Once the attack is performed, the optimized image closely resembles the ground truth training data.

Existing defenses against deep leakage attacks primarily focus on obfuscation strategies (including noise injection, gradient clipping, or pruning) to distort gradients and make data reconstruction more difficult \cite{zhu2019deepleakagegradients,wu2023concealingsensitivesamplesgradient,scheliga2021precodegenericmodel}. However, these approaches often suffer from two critical limitations: they degrade model utility and fail to prevent high-level semantic leakage. Even when reconstructions are noisy or distorted, an attacker can still infer the task being performed, the meaning of the class labels, and the general structure of the private data (Figure \ref{fig:spoofl}).

We propose a fundamental shift in defensive strategy: instead of merely obstructing the attack, we aim to exploit it. Rather than trying to prevent reconstruction entirely, we deliberately mislead the attacker into reconstructing plausible but entirely synthetic data that bears no immediately obvious relationship to the true data or learning task. By leveraging state-of-the-art distillation techniques and optimizing within the latent space of a powerful generative model trained on an unrelated external dataset, we guide the attacker toward false reconstructions that appear legitimate, but fail to reveal sensitive information.\cite{cazenavette2022datasetdistillationmatchingtraining}. This spoofing-based approach turns the attack against itself, ensuring the attacker believes they have succeeded, while actually learning nothing truthful about the real data. Our method achieves this without sacrificing significant model degradation, and opens a new paradigm in privacy-preserving Federated Learning: active deception over passive distortion. Our key contributions are:
\vspace{-10pt}
\begin{itemize}
\item \textbf{Redefining Deep Leakage defenses through spoofing-based quantification} – We shift the focus of DL defenses from passive obfuscation to active deception by introducing spoofing as a formal, quantifiable defense strategy. Traditional approaches like noise injection or gradient clipping may degrade image quality but still allow attackers to infer task semantics or class structure. In contrast, we propose deliberately guiding attackers toward reconstructing entirely synthetic, plausible-looking data that reveals nothing about the real training distribution. To support this, we introduce a new evaluation metric-Private Leakage Confidence (PLC), that measures how effectively a defense misleads an attacker from the private dataset. This redefinition of DL resilience in terms of adversarial misdirection establishes spoofing as a viable and measurable privacy-preserving paradigm in Federated Learning.

\item \textbf{A novel spoofing-based defense using Generator-optimized synthetic gradients} – We propose SpooFL, which optimizes the latent input of a generative model so that DL attacks recover fabricated samples instead of real training data. Unlike existing generative defenses that rely on private data, our approach leverages a generative network trained on an external dataset with no class overlap, ensuring that reconstructed images contain no meaningful information about the real training data, and mislead the attacker as to the nature of the FL task. We also demonstrate that SpooFL effectively lowers the number of communication rounds required to train an FL model due to its distilled nature.

\item \textbf{Comprehensive evaluation with established privacy and reconstruction metrics} – To rigorously assess data leakage, we introduce PLC  as a new metric, alongside a robust selection of recognized evaluation measures in the field to evaluate SpooFL against state-of-the-art defenses. This ensures a thorough, quantitative analysis of how effectively SpooFL misleads attackers while preserving model utility. It also provides a rigorous benchmarking regime to enable future works to explore this Spoofing based FL defense.  
\end{itemize}
\vspace{-15pt}

%% file: sec/2_lit.tex
\section{Related Work}
\label{sec:formatting}
\vspace{-7pt}
Federated Learning (FL) enables multiple clients to collaboratively train a shared model without exchanging raw data \cite{mcmahan2023communicationefficientlearningdeepnetworks}. Instead of sharing sensitive data, clients compute local updates and send them to a central server for aggregation, thereby enhancing privacy. By keeping raw data local and only transmitting model updates, FL reduces the risk of data exposure. In terms of update mechanisms, FedSGD transmits gradients after each batch, allowing for frequent updates but increasing communication costs. On the other hand, FedAvg reduces communication by performing multiple local updates before averaging the model weights \cite{DBLP:journals/corr/KonecnyMYRSB16}. However, this approach can introduce stale updates, especially when clients train on non-identically distributed (non-IID) data. Stale updates occur when local model updates become outdated before being aggregated into the global model, causing them to drift apart. This drift can lead to slower convergence or even model divergence \cite{rodio2024fedstaleleveragingstaleclient, karimireddy2021scaffoldstochasticcontrolledaveraging}. 

Despite the privacy guarantees offered by Federated Learning, the Deep Leakage from Gradients attack emerged in 2019 \cite{zhu2019deepleakagegradients}, which exploits shared gradients to reconstruct clients' private training data. This attack optimizes a randomly initialized dummy image to minimize the distance between the gradients it generates and the real gradients received by the attacker. Zhao et al. proposed ‘Improved Deep Leakage from Gradients (IDLG),’ an attack that extracts ground truth labels from the gradients of the last fully connected layer, offering faster convergence than DLG \cite{zhao2020idlgimproveddeepleakage}. Initially, these methods only worked on untrained networks, providing a false sense of security until the Inverting Gradients attack was introduced soon after \cite{geiping2020invertinggradientseasy}. This attack uses cosine similarity as the gradient distance function and Total Variation as a denoiser, enabling the recovery of images from gradients shared by trained networks. While Deep Leakage was primarily focused on the FedSGD configuration, Inverting Gradients was capable of extracting data from FedAvg by targeting the difference in model weights between updates. Most research on Deep Leakage defenses has focused on the FedSGD scenario, as it is easier to attack, but future attacks are likely to target FedAvg \cite{dimitrov2022dataleakagefederatedaveraging,zhao2022deepleakagemodelfederated}. Further advancements in Deep Leakage attacks often involve incorporating new regularization terms. For example, GradInversion introduces a Group Consistency Regularization term, which jointly optimizes multiple random seeds and enforces consistency by using the consensus image as a regularization constraint \cite{yin2021gradientsimagebatchrecovery}. Early Deep Leakage studies primarily focused on lower-resolution datasets \cite{huang2007labeled, netzer2011reading, krizhevsky2009learning, lecun1998mnist}, where gradients were computed using batch sizes of 1. However, recent research \cite{baglin2025fedladfederatedevaluationdeep,fowl2021robbing,geng2110towards} has demonstrated leakage on higher-resolution datasets, such as ImageNet \cite{deng2009imagenet} with larger batch sizes.

As Deep Leakage attacks continue to improve, demonstrating effectiveness on higher-resolution datasets, there is an increasing need for robust defenses to mitigate privacy risks in federated learning to achieve modern data regulation standards \cite{gdpr}. One well-established approach is Differential Privacy (DP), which ensures that the inclusion or exclusion of any individual data point has a minimal impact on the model’s output. Gradient Perturbation, a technique suggested since the original DL paper \cite{zhu2019deepleakagegradients}, aligns with DP principles by adding random noise to gradients (or pruning) before sharing them with the server. Over time, these techniques have advanced; for instance, the study “Securing Distributed SGD against Gradient Leakage Threats" \cite{wei2023securingdistributedsgdgradient} presents a method that adapts noise injection based on per-example gradient updates, enhancing privacy without significantly compromising model performance. Another common approach to defense focuses on modifying the training data itself rather than the shared gradients. One such strategy, “Defense by Concealing Sensitive Samples (DCS)" \cite{wu2023concealingsensitivesamplesgradient} optimizes a randomly initialized batch so that it appears visually distinct from the private training data while mimicking its gradient response. By disrupting the low entanglement among gradients within a batch, DCS hinders model inversion attacks that exploit such structure.  Similarly, PRECODE (PRivacy EnhanCing mODulE) \cite{scheliga2021precodegenericmodel} introduces stochasticity into gradient computations but does so by modifying the neural network itself rather than its training data. By integrating variational modeling into the architecture, PRECODE ensures that small changes in gradients do not correspond to deterministic updates, thereby thwarting gradient inversion. Subsequent enhancements, such as partial gradient perturbation and the Convolutional Variational Bottleneck (CVB), have further refined this approach, balancing privacy with model accuracy while minimizing computational overhead \cite{scheliga2023privacypreservingfederatedlearning}.

Building on these strategies, in 2025 Zhang et al. proposed FedADG (Federated Adversarial Data Generation) \cite{electronics14030406}, a GAN-based defense that replaces real gradients with those computed from synthetic pseudo-data, which are further perturbed with adaptive noise to obscure private information. This method aligns with PRECODE’s goal of disrupting gradient inversion but leverages generative modelling to introduce controlled obfuscation rather than architectural modifications. However, since the GAN is trained directly on private training data, sensitive information can still be leaked through shared labels and preserved feature distributions. In other words, the attacker can still infer the classes being learned and the nature of the data being used. There is also no guarantee that the ``synthetic" data may not end up being very similar, or indeed identical, to real data samples. Particularly if the underlying generative model suffers from mode collapse.

In contrast to these prior techniques, we propose approaching DL defence as a spoofing problem where we allow the attacker to recover the synthetic data, while making it appear to solve an unrelated problem to the true FL task. To demonstrate this idea we introduce SpooFL which optimizes a latent input to an unrelated generator producing synthetic samples that elicit gradients identical to those of the private training data. Because our generator is trained on a non-overlapping dataset, we ensure that private data and labels cannot be leaked. 

\vspace{-10pt}

%% file: sec/2.5_spoof.tex
\section{Spoofing Problem Definition}
\label{spoof_met}
\vspace{-7pt}
We first attempt to formally define the spoofing defense problem. Prior Deep Leakage defenses primarily aim to make the recovered images from attacks as noisy as possible. However, as advancements in Deep Leakage attacks improve their ability to denoise and bypass such defenses, these approaches become less effective. This suggests that current defense strategies are not future-proof, as they are only validated against existing attack methods and do not fundamentally prevent the recovery of private data. Therefore, it is crucial to reformulate how we measure successful Deep Leakage defenses to ensure that recovering private training data and labels is theoretically impossible. In this paper, we propose spoofing as the next direction in the field, offering a more robust solution to assure Federated Learning participants that their data remains secure.

In order to quantify the effectiveness of spoofing, we introduce a new metric: Private Leakage Confidence (PLC) which aims to determine whether the attacker can identify remnants of the true private class in its recovered data. PLC measures the confidence of a pretrained classifier that the reconstructed data  $x^*$ from the attack belongs to the original private class $y$. It is formally defined as:
\vspace{-5pt}
\begin{equation}
    PLC = \frac{C}{\mathcal{B}} \sum_{i=1}^{\mathcal{B}} f_{P}(x_i^*)\cdot \bf{y}_i ,
\end{equation}
\vspace{-1pt}
\noindent
where $C$ is the number of classes in the dataset, $\mathcal{B}$ is the batch size, $ f_{P}$ is the pretrained classifier on the private dataset and $\bf{y}$ is the one-hot encoding of the original private class. The mean confidence is normalized by multiplying by the number of classes $C$ to ensure that the metric remains independent of dataset size and class distribution. A better defense has a smaller PLC as that means it cannot correlate the recovered image to the correct class. It is important to ensure that the spoofed dataset contains no overlapping labels with the private training data. Any overlapping classes should be removed to maintain the integrity of the spoofing process, ensuring that the defense does not inadvertently train on data similar to the original dataset. This guarantees that the proposed metric accurately reflect the effectiveness of the spoofing strategy. 
\vspace{-10pt}

%% file: sec/3_method.tex
\section{Method}
\label{sec:method}
\begin{figure*}
    \centering
    \includegraphics[width=0.9\linewidth]{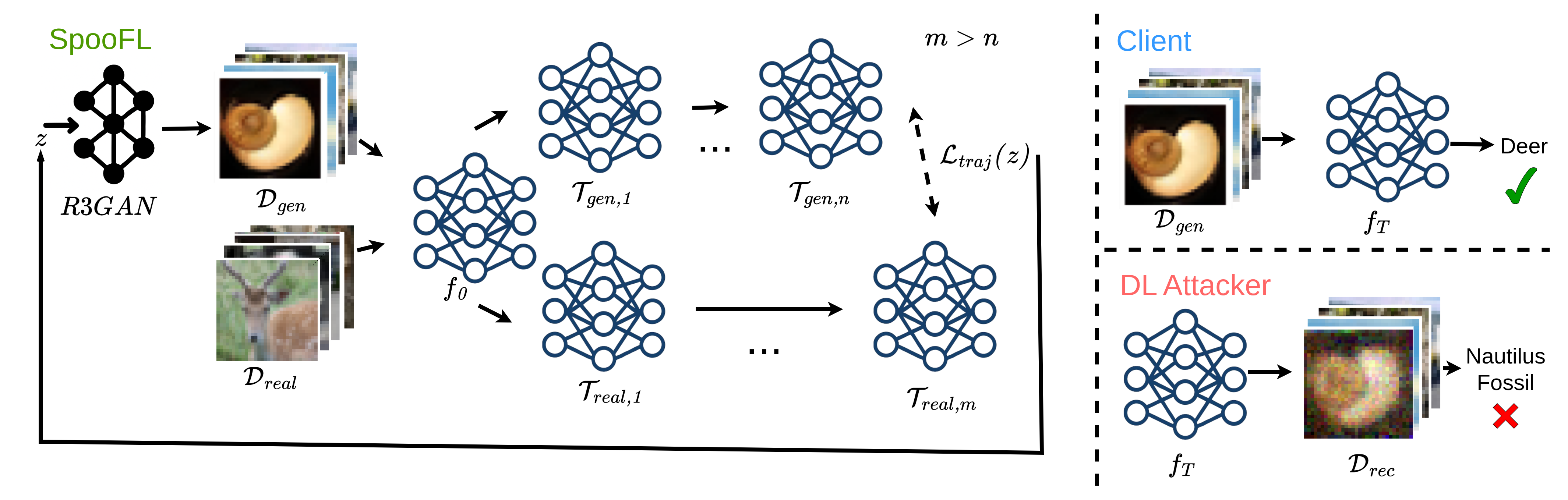}
    \vspace{-3pt}
    \caption{Overview of SpooFL. We optimize the latent input of an R3GAN using a model trajectory loss to distill a synthetic dataset. The resulting SpooFL dataset preserves task performance while effectively preventing deep leakage attacks by obfuscating private data.}
    \label{fig:traj}
    \vspace{-0.4cm}
\end{figure*}

\vspace{-5pt}
To illustrate the idea of Spoofing defenses in FL, our paper introduces SpooFL, a novel defense against Deep Leakage attacks in Federated Learning that misleads adversaries by ensuring reconstructed samples appear plausible but are entirely synthetic. Instead of adding noise to degrade reconstructions, SpooFL optimizes the latent vector of a state-of-the-art generative model trained on an external dataset to produce trajectory-matched synthetic images.  By leveraging an independent generative model with no class overlap, our method prevents feature and class leakage while maintaining model utility. This process is illustrated in Figure \ref{fig:traj}. SpooFL is implemented within the FEDLAD framework, allowing evaluation against state-of-the-art Deep Leakage attacks, defenses, and privacy metrics in a realistic Federated Learning setting \cite{baglin2025fedladfederatedevaluationdeep,DBLP:journals/corr/abs-2007-14390}.
\subsection{SpooFL}
\label{offline}
\vspace{-5pt}

SpooFL defends against Deep Leakage attacks by optimizing the latent input of a generative model to produce a spoofed dataset which misleads attackers into reconstructing entirely unrelated yet plausible images. This ensures no private data is exposed (including the nature of the target classes), while appearing to be a successful attack. To begin, the private training data must first be aligned with corresponding labels from an auxiliary dataset, referred to as the spoofing dataset, since SpooFL aims to generate images from this set. If the spoofing dataset contains overlapping classes, a blacklist $\mathcal{B}_{list}$ is curated to exclude these classes. Although we could select a random target spoof class, we can improve convergence and reduce model degradation by selecting the spoof class which is most similar to our sample. To this end, the spoofing label $\hat{y}_i $ for each private sample is estimated as:
\vspace{-5pt}
\begin{equation}\label{y_lab}
    \hat{y}_i = \arg\max_{c \notin \mathcal{B}_{list}} f_{S}(x_i)\cdot\bf{c}, \quad \forall i \in \{1, \dots, \mathcal{B}\},
\end{equation}
\vspace{-3pt}
\noindent
where $f_{S}$ is a pretrained classifier on the external dataset, $x_i$ represents a sample from the private training batch and $c$ is a class within the spoofing dataset. Once the spoofing label $\hat{y}_i$ is determined, we generate a spoofed dataset $\mathcal{D}_\text{gen}(z)$ using a generative model $G$. The generator takes as input a latent vector $z_{i}$ and the assigned spoofing label, producing a synthetic image:
\vspace{-3pt}
\begin{equation}\label{lateny}
    \mathcal{D}_\text{gen}(z) = G(\mathbf{z}_i| \hat{y}_i), \quad \forall i \in \{1, \dots, \mathcal{B}\},
\end{equation}
\vspace{-3pt}
\noindent
where, $G$ is a pretrained conditional generative model, such as a conditional GAN or a class-conditioned diffusion model, trained on the spoofing dataset. In our implementation, we use the generator of a pretrained conditional GAN, specifically the R3GAN model \cite{huang2025gandeadlonglive}, which achieves generative quality comparable to diffusion models while being significantly more computationally efficient. The latent vector $z_{i}$ serves as a learnable input that allows fine-grained control over the generated image, while conditioning $\hat{y}_i$ ensures that the generated sample belongs to the appropriate spoofing class.

Training-time optimization of latent vectors introduces significant computational overhead, making real-time (online) spoofed data generation challenging in certain settings. To address this, the SpooFL synthetic dataset is precomputed prior to model training. This spoofed dataset serves as a proxy for the private data, enabling downstream training without incurring any additional computation during the training phase. Our objective is to find a compact set of latent inputs such that the synthetic dataset produced by the generator (equation \ref{lateny}) induces a model training trajectory that matches that of a model trained on the entire real private data. 

We define the \textit{model trajectory} as a sequence of model states when a classifier is trained on the private dataset. Let \( \mathcal{T}_\text{real} \in \mathbb{R}^T \) denote the trajectory of a network \( f_\theta \) trained on the real private dataset \( \mathcal{D}_\text{real} \), over \( T \) epochs. Similarly, let \( \mathcal{T}_\text{gen}(z) \in \mathbb{R}^T \) be the trajectory when the classifier is trained on a synthetic dataset \( \mathcal{D}_\text{gen}(z)\), generated from a set of latent-label pairs \( z = \{(z_i, \hat{y}_i)\}_{i=1}^N \).

The \textit{model trajectory loss} is defined as
\vspace{-5pt}
\begin{equation}
\mathcal{L}_{\text{traj}}(z) = \frac{1}{T} \sum_{t=1}^{T} \left\| \mathcal{T}_\text{gen}^{(t+n)}(z) - \mathcal{T}_\text{real}^{(t+m)} \right\|^2,
\end{equation}
\vspace{-3pt}
\noindent
where \( \mathcal{T}_\text{gen}^{(t)}(z) \) denotes the model trajectory of the classifier at epoch \( t \) when trained on \( \mathcal{D}_\text{gen}(z) \), and \( \mathcal{T}_\text{real}^{(t)} \) is the corresponding value from the real dataset trajectory. $n$ and $m$ are offsets that provide flexibility in aligning and comparing the trajectories at different training steps. This loss extends the idea of gradient matching from individual items (during FL attacks) to the dataset as a whole.

Our goal is to find latent inputs \( z \) such that the generated data induces a similar training behavior to the real data:
\vspace{-4pt}
\begin{equation}
\hat{z} = \arg\min_z \mathcal{L}_{\text{traj}}(z).
\vspace{-4pt}
\end{equation}
\noindent
Here, $\hat{z}$ is the optimized latent vector after 200 iterations of AdamW with a learning rate of 0.1. Jointly optimizing latent representations for every example in a large dataset using this approach can be computationally challenging. To overcome this, we synthesize a much smaller set of optimized latent vectors (corresponding to synthetic (spoofed) training examples) that approximate the learning efficacy of the full dataset. In practice, this enables us to reduce private datasets comprising hundreds of thousands of samples to as few as 100 spoofed images. To prevent overfitting and improve model accuracy, the distillation process can be repeated until a sufficiently diverse and informative spoofed dataset is constructed.
\vspace{-7pt}
\subsection{Theoretical Intuition for Spoofing Defenses}
\vspace{-5pt}
 At first glance, it may seem counter-intuitive that a dataset can be generated to effectively train a model for a task; when none of the samples in the dataset belong to any of the classes involved. However, it has long been known that, due to the high-dimensional and often piecewise-linear nature of deep learning models, even minor perturbations to an input \( \mathbf{x} \) can cause significant shifts in classification. Goodfellow et. al found that adversarial examples \cite{goodfellow2015explainingharnessingadversarialexamples} can exploit this by introducing a small perturbation \( \boldsymbol{\eta} \) such that the perturbed input \( \mathbf{x}' = \mathbf{x} + \boldsymbol{\eta} \) remains visually indistinguishable but leads to a different prediction. This occurs because the change in the model's output is approximately linear:  
 \vspace{-3pt}
\begin{equation}
    g(\mathbf{x} + \boldsymbol{\eta}) \approx g(\mathbf{x}) + \nabla g(\mathbf{x})^\top \boldsymbol{\eta}.
\end{equation}
 \vspace{-3pt}
When \( \boldsymbol{\eta} \) is aligned with \( \nabla g(\mathbf{x}) \), even small perturbations can accumulate and produce large shifts in classification. If this shift is large enough to cross the model’s decision boundary, it can flip the predicted class, despite the perturbation being imperceptible to humans. This observation highlights how neural networks exhibit excessive linearity in high-dimensional spaces, making them highly sensitive to structured perturbations that drastically alter predictions. 

SpooFL (and spoofing defenses more generally) exploit this same sensitivity, but in a constructive way. Rather than introducing small perturbations to fool a classifier, we deliberately optimize synthetic inputs so that their responses align with those that would have been induced by the private data. In other words, spoofed samples need not resemble the private data visually — they only need to reproduce the same network behavior when passed through the model. This distinction is critical: adversarial examples aim to preserve appearance while altering predictions, whereas spoofed data aims to alter appearance while preserving model training trajectories. As a result, an external observer attempting gradient inversion will reconstruct images that appear coherent but belong entirely to an unrelated domain, while the federated model continues to learn the true private task. This dual effect (preserving utility while misleading attackers) arises directly from the over-parameterization and gradient sensitivity of deep models, which make trajectory-aligned spoofing feasible in practice.

%% file: sec/4_eval.tex
\vspace{-5pt}
\section{Evaluation}
\label{sec:eval}
\vspace{-3pt}
\textbf{FL Setup.} We evaluate SpooFL in standard federated image classification settings, matching the protocols of prior works. All experiments adopt the FedAvg protocol, where clients perform several local steps before transmitting model updates to the server. We consider an honest-but-curious server threat model in which the server (or a passive eavesdropper) attempts to reconstruct private training samples from intercepted updates. Models include ResNet-18 \cite{7780459}, VGG-11 \cite{simonyan2015deepconvolutionalnetworkslargescale}, and a lightweight custom Convnet\cite{wu2023concealingsensitivesamplesgradient}. Training is performed with a cross-entropy loss and the SGD optimizer (learning rate 0.01). Experiments run on a single 24GB NVIDIA RTX 3090 GPU (CUDA 12.9), running Ubuntu 20.04. Evaluations are conducted on two widely used benchmarks: (1) CIFAR-10, a 10-class dataset of 32×32 images \cite{cifar10}, and (2) STL-10, a 10-class dataset of higher-resolution 96×96 images \cite{stl10}. These datasets provide complementary scales for assessing both reconstruction difficulty and defense robustness. SpooFL employs R3GAN as the generative backbone, trained on an external dataset (ImageNet) \cite{huang2025gandeadlonglive}. By default, reconstructions are carried out using the SME attack \cite{zhu2023surrogatemodelextensionsme} with 30,000 optimization iterations.

\textbf{Evaluation Metrics.} To comprehensively assess privacy and utility, we report both classical reconstruction metrics and our newly introduced spoofing-oriented privacy metric: 
\vspace{-18pt}
\begin{itemize}[topsep=2pt, itemsep=2pt, parsep=0pt]
    \item \textbf{Structural Similarity Index (SSIM ↓):} Measures pixel-wise structural overlap between reconstructions and ground truth.
    \item \textbf{Feature Mean Squared Error (FMSE ↑):} Computes embedding-space divergence, where higher values indicate reconstructions are dissimilar to private data.
    \item \textbf{Peak Signal-to-Noise Ratio (PSNR ↓):} Reports reconstruction fidelity; lower values imply poorer alignment with the private sample.
    \item \textbf{LPIPS (↑):} Learned perceptual similarity in feature space; higher values reflect greater divergence from the ground truth(VGG backbone).
    \item \textbf{Private Leakage Confidence (PLC ↓):} Our proposed privacy metric quantifying the extent to which reconstructed images are confidently assigned to the correct private label distribution. Lower PLC indicates stronger spoofing.
    \item \textbf{Model Accuracy (↑):} Ensures training on spoofed data preserves downstream classification performance.
    \item \textbf{Relative Execution Time (1.0 baseline):} Measures runtime overhead compared to training without any defense.
\end{itemize}

\begin{table}[H]
\caption{Comparison of different defense methods (noise injection, clipping, and compression) on ResNet-18 model accuracy and deep leakage reconstruction quality.}
\centering
\resizebox{0.45\textwidth}{!}{%
\begin{tabular}{c|c|c|c}
\hline
\textbf{Defense Method} & \textbf{Defense Parameter} & \textbf{Accuracy (\%)} $\uparrow$ & \textbf{SSIM} $\downarrow$ \\
\hline
None & -- & 75.43 & 0.587 \\
\hline
 & $1.0 \times 10^{-3}$ & \textbf{72.56} & 0.335 \\
Noise & $2.5 \times 10^{-3}$ & 63.33 & 0.273 \\
 Injection & $5.0 \times 10^{-3}$ & 51.99 & 0.147 \\
 & $1.0 \times 10^{-2}$ & 35.86 & \textbf{0.138} \\
\hline
 & 20.0 & \textbf{65.74} & 0.460 \\
Clipping & 15.0 & 54.74 & 0.421 \\
& 10.0 & 46.70 & 0.375 \\
& 5.0  & 26.80 & \textbf{0.196} \\
\hline
 & 90.0   &\textbf{ 57.82 }& 0.484 \\
Compression & 92.5 & 57.20 & 0.435 \\
 & 95.0   & 56.62    & 0.321 \\
 & 97.5 & 40.06 & \textbf{0.093} \\
\hline
\end{tabular}
}

\label{tab:defenses_vs_ssim}

\end{table}
\textbf{Defense Implementations.} In all cases, defense parameters are chosen to balance privacy protection with the practical need to preserve model accuracy. Using overly aggressive settings (such as very high noise scales) could trivially prevent reconstruction but at the cost of rendering the model unusable (see Table \ref{tab:defenses_vs_ssim}). To ensure meaningful comparison, we therefore adopt moderate configurations: (1) Noise: additive perturbations drawn from a Gaussian Distribution with $\sigma$ fixed at $2.5 \times 10^{-3}$, providing obfuscation without destabilizing training; (2) Clipping: bounding update magnitudes with a maximum value of 20; (3) Compression: sparsifying updates at a 95\% rate; (4) DCS: pseudo-batch optimization following the original implementation \cite{wu2023concealingsensitivesamplesgradient}; (5) FedADG: replacement of client updates with gradients derived from GAN-generated samples \cite{electronics14030406}; and (6) SpooFL (ours): a compact distilled dataset  is generated ahead of time to provide efficient spoofing defense.

\textbf{Attack Baselines.} To evaluate the robustness of our defense, SpooFL, we consider three representative deep leakage attacks that target private data in federated learning: 
\begin{itemize}[topsep=2pt, itemsep=2pt, parsep=0pt]
   \item  \textbf{GradInversion (GI):} Improves stability and visual fidelity by minimizing an $\ell_2$ gradient matching loss, optimized with Adam. It further employs a consensus regularization loss by jointly optimizing multiple random seeds \cite{yin2021gradientsimagebatchrecovery}. 
 
  \item \textbf{Data Leakage in Federated Averaging (DLF):} Combines a simulation-based gradient matching loss ($L_{sim}$), which explicitly simulates the client’s local training steps, with an epoch-order invariant prior ($L_{inv}$) that regularizes reconstructions across epochs despite random batch order\cite{dimitrov2022dataleakagefederatedaveraging}. 
  
   \item \textbf{Surrogate Model Extension (SME):} Builds surrogate states by linearly combining pre- and post-update weights to approximate intermediate hidden models. Reconstruction is performed by minimizing a gradient matching loss based on cosine similarity, optimized using Adam \cite{zhu2023surrogatemodelextensionsme}.
\end{itemize}
\subsection{Defense Performance}
Table \ref{tab:deep_leakage_defenses} presents a comparison of SpooFL against state-of-the-art defenses on a trained ResNet18 model on CIFAR-10 and STL-10. Aggregated across both datasets SpooFL achieves the lowest Private Leakage Confidence (PLC), indicating that attacker reconstructions are better separated from the true private distribution. Importantly, this privacy gain does not come at the expense of model utility: SpooFL maintains competitive or superior accuracy compared to other defenses. Although aggressive defenses such as Compression or DCS come close (and occasionally surpass) SpooFL in PLC and reconstruction metrics, they achieve this by degrading FL performance far more severely. In addition, SpooFL introduces no additional runtime cost, with a Relative Execution Time (RET) of 1.0, whereas alternatives like DCS incur more than 4× overhead. FedADG achieves strong defense across most metrics and preserves high model accuracy; however, because its generative model is trained on the private dataset, the PLC remains elevated. Taken together, these results highlight that SpooFL provides a highly favorable privacy–utility tradeoff, simultaneously reducing PLC, sustaining accuracy, and avoiding costly training slowdowns.

\begin{table*}[h]
\centering
\caption{ResNet18: Performance comparison of SpooFL and state-of-the-art Deep Leakage defenses on CIFAR-10 and STL (batch size of 8), evaluating privacy, execution time, and model accuracy (SME attack). Results for FedADG were omitted for STL due to the lack of an effective generative network.}
\resizebox{1\textwidth}{!}{%
\renewcommand{\arraystretch}{1.3}
\begin{tabular}{ c c c c c c c c c }
\hline
 \textbf{Dataset} & \textbf{Defense Method} & \textbf{SSIM} $\downarrow$ & \textbf{FMSE} $\uparrow$  & \textbf{PSNR} $\downarrow$  & \textbf{LPIPS} $\uparrow$  & \textbf{PLC} $\downarrow$ & \textbf{Relative Execution Time} $\downarrow$ & \textbf{Model Accuracy(\%)} $\uparrow$  \\ \hline
 & Ground Truth & 1.000 & 0.000  & -- & 0.000 & 3.843  & 1.00  & 73  \\
 & None  & 0.609 &1.083  &19.127  &0.323  &3.832  & 1.00 & 73  \\  
 \cline{2-9}
 & Gaussian Noise  & 0.273 &2.215  &13.509  &0.515  &2.353  &1.02  & 63 \\ 
 CIFAR- & Clipping  & 0.460  & 1.339  & 16.777  &0.515  &3.126  &1.03 &\textbf{66}  \\               
 10 & Compression  &0.321  &1.608  &13.882  &0.462  &3.067  & 1.20 & 56 \\  
 & DCS  & 0.155  &2.645  & 12.223  & \textbf{0.586}  &1.822  &4.18  &60  \\ 
  & FedADG  & 0.182 & 1.562  &12.332  &0.492  &2.651  & 1.30 & 65 \\ 
  & SpooFL  & \textbf{0.139} &\textbf{6.513}  &\textbf{11.507}  &0.576 & \textbf{1.689}  & \textbf{1.00 } & 63  \\ 
 \hline
& Ground Truth &1.000  &0.000   &--   &0.000& 3.745  &1.00   &  56  \\
& None  & 0.486  &0.114   &17.008   &0.373   &3.478   &1.00   &   56 \\  
 \cline{2-9}
& Gaussian Noise  &0.190   &1.345   &13.154& 0.629   &1.700   &1.02      &48    \\ 
 STL & Clipping  & 0.297 & 2.022 & 13.832   &0.558   &  1.765  &1.03     &  \textbf{50}  \\               
 & Compression  & 0.159  &1.392   &12.457   &0.514   &\textbf{1.493 }  & 1.20  &  44  \\  
 & DCS  & 0.154  & 2.375   & \textbf{12.137 }  & 0.593  & 1.657   & 4.18   & 45    \\ 
 & FedADG  & --  & --  & --  & --  & --  & --  & --   \\ 
& SpooFL  & \textbf{0.140}   &\textbf{ 2.948 }  &12.551   &\textbf{0.636  } &1.699   &\textbf{1.00   }& 48    \\ 
\hline
\end{tabular}%
}
\label{tab:deep_leakage_defenses}
\end{table*}

\begin{figure}
\centering

\begin{tikzpicture}
  \begin{axis}[
    xlabel={Dataset Size},
    ylabel={Model Accuracy (\%)},
    xmin=0, xmax=20000,
    ymin=0, ymax=80,
    grid=both,
    width=0.6\linewidth,  % half page width
    height=0.45\linewidth, % proportional height (adjust as needed)
    mark size=2pt,
    legend pos=south east,
    legend style={font=\footnotesize},
    ]
    \addplot[
      color=blue,
      mark=*,
      ]
      coordinates {
        (20000,63)
        (15000,62)
        (10000,59)
        (5000,56)
        (2500,54)
        (1000,48)
        (500,38)
        (250,33)
        (100,20)
        (50,17)
        (0,10)
        
      }; \addlegendentry{SpooFL}
          \addplot[
      color=orange,
      mark=*,
      ]
      coordinates {
        (20000, 67)      
        (15000,63)
        (10000,57)
        (5000,48)
        (2500,34)
        (1000,29)
        (500,20)
        (250,17)
        (100,10)
        (50,10)
        (0,10)
        
      }; \addlegendentry{CIFAR10}
      
  \end{axis}
  
\end{tikzpicture}
\vspace{-5pt}
\caption{Dataset size versus ResNet-18 model accuracy on CIFAR-10 and SpooFL. SpooFL achieves competitive accuracy with significantly fewer training samples, highlighting its efficiency in reducing communication and computation overhead.}
\label{size}
\vspace{-20pt}
\end{figure}
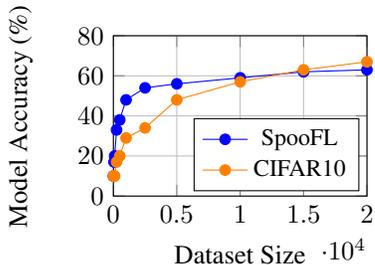

Figure \ref{size} reports model accuracy as a function of dataset size when training with the SpooFL dataset compared to the full CIFAR-10 training set. As expected, accuracy improves for both settings as more data is used; however, CIFAR-10 begins to outperform once the dataset size grows large. The key observation is that SpooFL attains competitive accuracy even at substantially smaller dataset sizes. This demonstrates that if one is willing to sacrifice a small amount of performance, high utility can be achieved with a significantly reduced dataset via SpooFL. In practice, this reduction translates directly into fewer communication rounds and lower resource requirements in federated training, offering strong efficiency gains while still maintaining strong defense properties.

\begin{figure}[t]
\vspace{-5pt}
    \centering
    \includegraphics[width=1\linewidth]{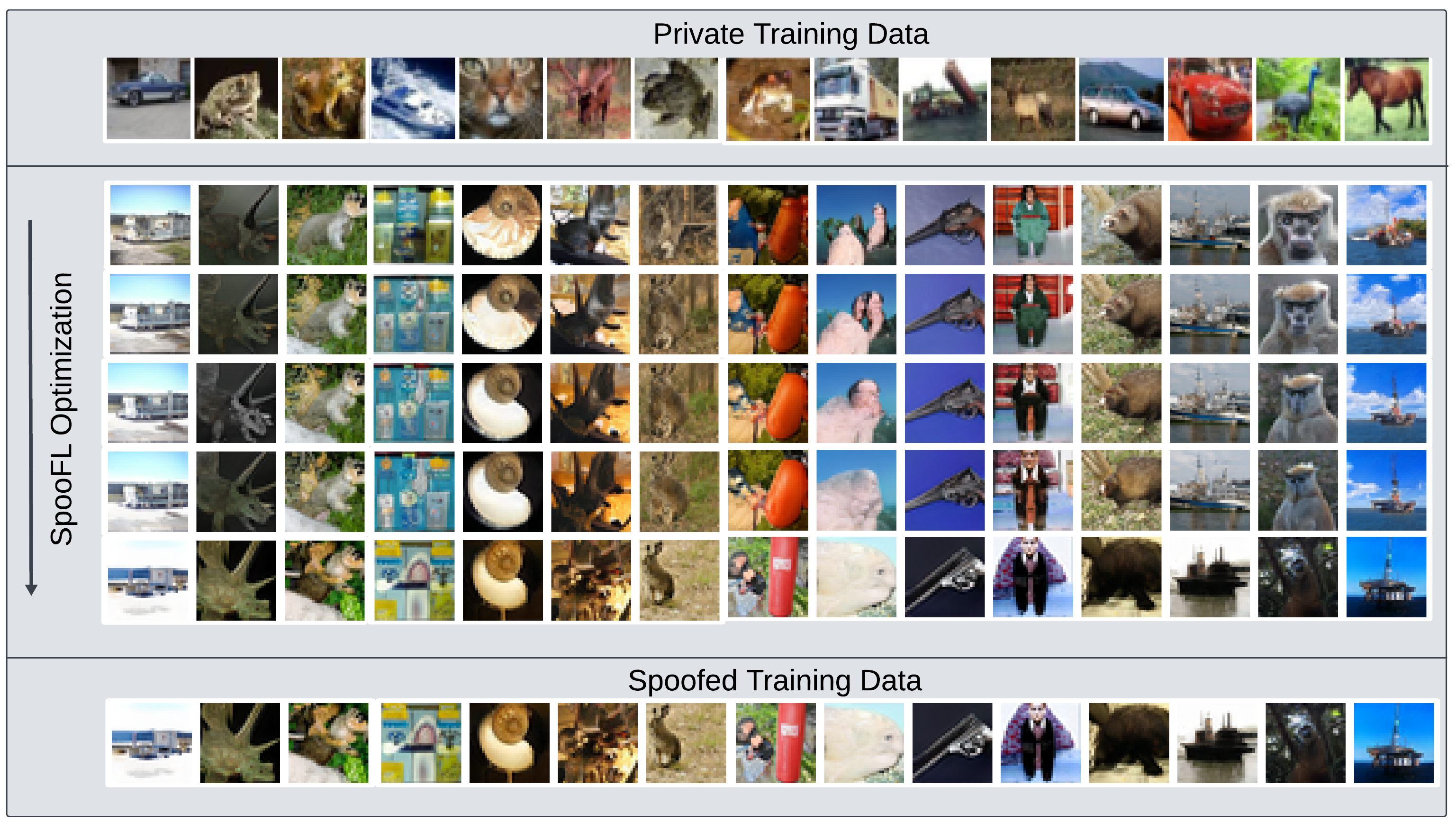}
    \vspace{-5pt}
    \caption{Visualization of the SpooFL distillation process. Synthetic images are iteratively optimized to match the model trajectory of the private dataset. The final distilled images converge to representations that follow the same training dynamics as the original private data, without revealing sensitive information. More examples given in the supplementary material.}
    \label{fig:placeholder}
    \vspace{-10pt}
\end{figure}

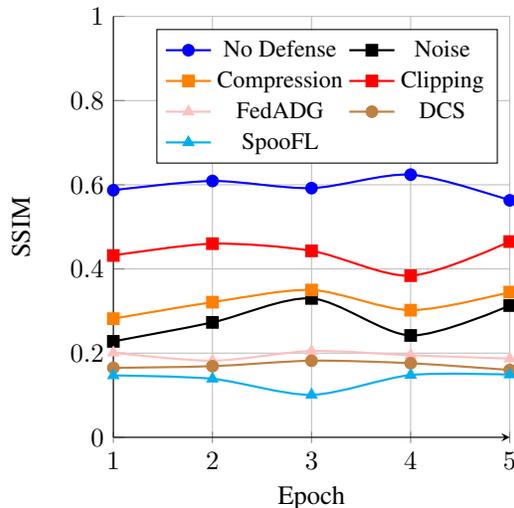
\begin{figure}[t]
\centering
\begin{tikzpicture}

\begin{axis}[
    width=0.83\linewidth,  % half page width
    height=0.87\linewidth, % proportional height (adjust as needed)
    xlabel={Epoch},
    ylabel={SSIM},
    xmin=1, xmax=5,
    ymin=0, ymax=1,
    xtick={0,1,2,3,4,5},
    ytick={0,0.2,0.4,0.6,0.8,1.0},
    legend style={at={(0.1,-0.2)}, anchor=north, legend columns=2,font=\footnotesize}, column sep=0.01pt,
    legend pos=north east,
    axis y line*=left,
    axis x line=bottom,
    grid=major,
]

% SME SSIM (blue, smooth)
\addplot[blue, thick, mark=*, smooth] coordinates {
    (1,0.587)
    (2,0.609)
    (3,0.592)
    (4,0.624)
    (5,0.563)
};
\addlegendentry{No Defense}

\addplot[black, thick, mark=square*, smooth] coordinates {
    (1,0.228)
    (2,0.273)
    (3,0.330)
    (4,0.242)
    (5,0.313)
};
\addlegendentry{Noise}

% Ours SSIM (red, smooth)
\addplot[orange, thick, mark=square*, smooth] coordinates {
    (1,0.282)
    (2,0.321)
    (3,0.350)
    (4,0.302)
    (5,0.345)
};
\addlegendentry{Compression}
\addplot[red, thick, mark=square*,smooth] coordinates {
    (1,0.432)
    (2,0.460)
    (3,0.443)
    (4,0.384)
    (5,0.465)
};
\addlegendentry{Clipping}

\addplot[pink, thick, mark=triangle*,smooth] coordinates {
    (1,0.201)
    (2,0.182)
    (3,0.205)
    (4,0.195)
    (5,0.187)
};
\addlegendentry{FedADG}

\addplot[brown, thick, mark=*,smooth] coordinates {
    (1,0.165)
    (2,0.169)
    (3,0.182)
    (4,0.176)
    (5,0.16)
};
\addlegendentry{DCS}

\addplot[cyan, thick, mark=triangle*,smooth] coordinates {
    (1,0.147)
    (2,0.139)
    (3,0.101)
    (4,0.148)
    (5,0.149)
};
\addlegendentry{SpooFL}

\end{axis}

\end{tikzpicture}
\vspace{-10pt}
\caption{SSIM evolution over training epochs under different defense methods. SpooFL consistently maintains low SSIM values, indicating strong resistance to reconstruction throughout training.}
\label{training}
\vspace{-15pt}
\end{figure}

\begin{table}[t]
\centering
\caption{Effectiveness of SpooFL defense against GradInversion, DLF, and SME attacks on CIFAR-10.}
\resizebox{1\linewidth}{!}{%
\renewcommand{\arraystretch}{1.3}
\begin{tabular}{ c c c c c  c}
\hline
 \textbf{Attack Method} & \textbf{SSIM} $\downarrow$ & \textbf{FMSE} $\uparrow$  & \textbf{PSNR} $\downarrow$  & \textbf{LPIPS} $\uparrow$  & \textbf{PLC} $\downarrow$  \\ \hline
GradInversion   & 0.078 &3.080  &6.564  &0.578     &0.864  \\
GradInversion (SpooFL)  & 0.023 & 7.921  & 4.788     & 0.615  & 0.165  \\
\hline
DLF  &0.253  &1.615  &9.573  &0.425  &2.392      \\
DLF(SpooFL)  &0.125  &7.234  &8.254     &0.629  &1.891  \\ 
\hline
SME  & 0.609 &1.0828  &19.127  &0.323 &3.832     \\ 
SME (SpooFL)  &0.139 & 6.512 &11.507 &0.676  &1.689      \\ 

\hline
\end{tabular}%
}

\label{tab:deep_leakage_defenses3}
\vspace{-7pt}
\end{table}
Table \ref{tab:deep_leakage_defenses3} examines the robustness of SpooFL under different reconstruction attacks. We test against three widely used attack methods (GradInversion, DLF, and SME) and report privacy leakage across both CIFAR-10 and STL-10. In all cases, SpooFL significantly reduces the PLC compared to having no defense, substantially reducing the attacker’s ability to infer private information. For example, applying SpooFL lowers GradInversion's PLC by more than half compared to the undefended baseline, while accuracy remains competitive. This indicates that the spoofing mechanism generalizes effectively across attack strategies, not just the default SME setup. Table \ref{tab:deep_leakage_defenses2} evaluates SpooFL across multiple model architectures, including ResNet-18, VGG-11, and a lightweight  Convnet. While absolute accuracy varies with architecture, as expected, SpooFL preserves model utility while maintaining strong protection. Shallower networks tend to be more sensitive to obfuscation methods, whereas deeper models are generally better at balancing the utility–privacy trade-off. Together, Tables  \ref{tab:deep_leakage_defenses3} and \ref{tab:deep_leakage_defenses2} demonstrate that SpooFL is robust both to variations in the attacker’s strategy and to differences in the target model.

\begin{table}[t]
\centering
\caption{Cross-architecture evaluation of SpooFL on ResNet-18, VGG-11, and a custom Convnet for CIFAR10}
\resizebox{1\linewidth}{!}{%
\renewcommand{\arraystretch}{1.3}
\begin{tabular}{c c c c c  c}
\hline
 \textbf{Model architecture} &\textbf{Defense Method} & \textbf{SSIM} $\downarrow$  & \textbf{PSNR} $\downarrow$  & \textbf{LPIPS} $\uparrow$  & \textbf{PLC} $\downarrow$  \\ \hline
 &None & 0.621 & 18.314   & 0.313     & 3.162 \\
 & Gaussian Noise& 0.443 &  16.402       &  0.394 & 2.087  \\
 
Custom Convnet   & Clipping &0.421   &15.565 &0.395  &1.987      \\
 & Compression& 0.512 &16.527     &0.317 &2.435  \\ 
&SpooFL &\textbf{0.155}  &\textbf{12.522}   &\textbf{0.624} & \textbf{1.030}     \\ 
\hline
&None    &0.472 & 15.524 &0.406  &2.805      \\ 
&Gaussian Noise & 0.285   & 13.045 &0.592  &1.831      \\ 
VGG11 &Clipping &0.243   &12.876 &0.615  &1.673      \\ 
& Compression & 0.182  &12.563 & \textbf{0.620}  &\textbf{1.537 }     \\ 
& SpooFL& \textbf{0.152}  &\textbf{12.545} &0.603& 1.561     \\ 
\hline
&None &0.609   & 19.127 & 0.323  &3.832      \\ 
&Gaussian Noise & 0.273  &13.509 &0.515  &2.353      \\ 
ResNet18 &Clipping &0.460   &16.777 &0.515  &3.126      \\ 
& Compression & 0.321  & 13.882 &0.462  &3.067      \\ 
& SpooFL& \textbf{0.139}  &\textbf{11.507} &\textbf{0.576}  &\textbf{1.689}      \\ 
\hline
\end{tabular}%
}

\label{tab:deep_leakage_defenses2}
\vspace{-20pt}
\end{table}

Figure \ref{fig:placeholder} shows qualitative intermediate examples of the SpooFL distillation process where samples from a range of true classes, are distilled into equivalent examples of unrelated spoofed classes. Figure \ref{training} illustrates how defenses behave throughout the training process. These qualitative trends reinforce the quantitative results in Table \ref{tab:deep_leakage_defenses}, showing that SpooFL remains effective across training, rather than providing only transient protection.

%% file: sec/5_conc.tex
\vspace{-10pt}
\section{Conclusion}
\label{sec:conc}
\vspace{-7pt}
In this work, we explored spoofing as a novel approach to defending against Deep Leakage (DL) attacks in Federated Learning (FL). Unlike traditional defenses that focus on obfuscation or noise injection, spoofing actively misleads attackers by generating plausible yet entirely synthetic reconstructions. This not only prevents meaningful data leakage but also disrupts an attacker's understanding of the task being solved. As a concrete realization of this concept, we introduced SpooFL, which leverages a state-of-the-art conditional generative model trained on an external dataset to achieve this deception while maintaining model utility. To rigorously assess our approach, we introduced a new privacy metric, Private Leakage Confidence (PLC), which quantify the extent to which reconstructed samples resemble private data. Through extensive evaluation, we demonstrated that SpooFL effectively deceives attackers, outperforming conventional defenses in preventing data exposure.

As Deep Leakage attacks evolve, developing adaptive and scalable privacy-preserving techniques will be crucial for secure FL deployments. One limitation of SpooFL is that the spoofed dataset is tailored to the specific network used during distillation and may not generalize to other architectures without re-distillation. Future work should explore improving generative models for more efficient spoofing and reducing dependence on network-specific datasets.